\documentclass{jfm}
\usepackage{graphicx}
\usepackage{epstopdf, epsfig}
\usepackage{comment}
\usepackage{xcolor}
\usepackage{amsmath}

\shorttitle{Cavity closure dynamics}
\shortauthor{T. Kroeze, D. Fernandez Rivas and  M.A. Quetzeri-Santiago}

\title{Gas density influences the transition from capillary collapse to surface seal in microfluidic jet impacts on deep pools}

\author{Thijmen B. Kroeze\aff{1},
  David Fernandez Rivas \aff{1},
 \and  Miguel A. Quetzeri-Santiago\aff{2}
 \corresp{\email{mquetzeri@materiales.unam.mx}}}

\affiliation{\aff{1}Mesoscale Chemical Systems Group, MESA+ Institute and Faculty of Science and Technology, University of Twente, P.O. Box 217, 7500AE Enschede, The Netherlands \\
\aff{2}Instituto de Investigaciones en Materiales, Universidad Nacional Autónoma de México, Cd. Universitaria, 04510 Mexico City, Mexico}
\begin{document}

\maketitle

\begin{abstract}
Studies of liquid jet impacts onto a deep liquid pool are of great significance for a multitude of engineering and environmental applications. During jet impact, the free surface of the pool deforms and a cavity is generated. Simultaneously, the free surface of the cavity extends radially outward and forms a rim. Eventually the cavity collapses by means of gas inertia and surface tension. In this work we study numerically such cavity collapse, under different impact velocities and ambient gas density conditions. An axisymmetric numerical model, based on the volume of fluid method is constructed in Basilisk C. This model is validated by qualitative and quantitative comparison with theory and experiments, in a parameter range that has not been previously explored. Our results show two distinct regimes in the cavity collapse mechanism. By considering forces pulling along the interface, we derive scaling arguments for the time of closure and maximum radius of the cavity, based on the Weber number. For jets with uniform constant velocity from tip to tail and $We \leq 150$ the cavity closure is capillary dominated and happens below the surface (deep seal). In contrast, for $We \geq 200$ the cavity closure happens above the surface (surface seal) and is dominated by the gas entrainment and the pressure gradient that it causes. Our results provide information for understanding pollutant transport during droplet impacts on large bodies of water, and other engineering applications, like additive manufacturing, lithography and needle-free injections. 
\end{abstract}

\section{Introduction}

The pioneering work of \cite{worthington1908} displaying and describing liquid impacts onto pools initiated a century-long interest into characterising such impact phenomena. Understanding the intricacies of these events is relevant for a broad spectrum of situations in nature; the noise of rain \citep{prosperetti1989} or the scent of earth after rain on a hot day \citep{joung2015}, as well as in technology such as in inkjet printing \citep{vanderbos2014} or spray atomisation \citep{panao2005}. Of particular interest has been the air entrainment, cavity formation and collapse \citep{lee1997cavity,truscott2014water,Eshraghi2020,eggers2007theory, deka2018dynamics}. In general, the cavity formation in pools begins when the jet impacts the free surface of the target and deflects its surface. This deflection occurs just before coalescence of the jet and the bath, as the local gas pressure builds up by the approaching liquid jet \citep{bouwhuis2015}. Upon coalescence the inertia of the high-speed jet dominates the deflection and a hemispherical cavity is formed by the head of the jet \citep{speirs2018}. Due to its radial expansion, the cavity has lamella shooting radially outwards, extending the walls of the cavity. The remainder of the jet impacts the base of the newly formed cavity and extends it primarily in the direction of travel, producing a slender cavity \citep{bouwhuis2015}. In this process the kinetic energy of the jet is converted to potential surface energy and heat due to dissipation \citep{speirs2018}.

There is, however, a stark disparity in the amount of work done on projectiles in the millimetre regime \citep{engel1966,pumphrey1990,rein1993,oguz1995,yarin2006,bartolo2006, aristoff2009water, zhang2012evolution, agbaglah2014growth, truscott2014water, fudge2021dipping} compared to projectiles in the micrometer regime \citep{bouwhuis2016,speirs2018,miguel2021,quetzeri2023cavity}. For cavities generated by projectiles in the millimetre range the collapse can be mainly attributed to the hydrostatic pressure \citep{oguz1995}. This implies that the Bond number ($Bo = \rho_0 g D^2/ \gamma$) is greater than one (where $\rho_0$ and $\gamma$ are the density and surface tension of the liquid, $D$ is a relevant length scale and g the acceleration due to gravity). The domain of interest for our work is in the micrometer regime, where collapse is driven by surface tension forces as $Bo \sim O(10^{-3})$. This regime is relevant in emerging technologies as 3D printing \citep{antkowiak2011}, spray painting \citep{Herczynski2011}, extreme ultraviolet lithography \citep{klein2015}, environmental aspects \citep{speirs2023capture} and needle free injection methods \citep{Rodriquez2016,Galvez2020,van2023microfluidic}.

In this work, we looked into the dynamics of a high-speed microfluidic jet penetrating a pool. These jets are comparable in size and momentum produced in needle free applications \citep{schoppink2022jet}. A validation process was done through both qualitative and quantitative comparisons with other numerical results, experiments and theoretical predictions. In addition we quantified the cavity profile and closure time of the cavity as a function of relevant fluid parameters. Our numerical strategy provides the opportunity to examine a broad parameter space unconstrained from experimental limitations.

\section{Methodology}

\subsection*{Experimental details}

A transparent cubic bath made of acrylic with dimensions of $5\times 10\times 20$ cm, was filled with water. High-speed jets were generated from a thermocavitation process and directed to impact a water pool. The setup is similar to the ones used in refs. \citep{miguel2021, quetzeri2023cavity}. The thermocavitation process occurs inside a glass microfluidic chip filled a Direct Red 81 solution in water at  0.5 wt. $\%$. In thermocavitation, a expanding bubble is created at the base of the chip, due to the energy transfer to the liquid from a continuous wave laser. The expanding bubble pushes the liquid that is in front of it generating the jet \citep{oyarte2020microfluidics}. The jet velocity $U_j$ and diameter $R_j$ in these experiments ranged from 10 m/s to 40 m/s, and 25-100 $\mu$m respectively. The surface tension of water is $\gamma = 0.072$ N/m, its density $\rho_0 = 1000$ kg/m$^3$ and its viscosity $\mu = 1$ cP. Thus, the Weber number $We = \rho_0 U_0^2 R_j /\gamma$, and the Reynolds numbers  $Re = \rho_0 U_0 R_j / \mu$ range between 35-1333 and 500-4000, respectively.
The processes of bubble generation, jet ejection and impact on the liquid droplet were recorded with a Photron Fastcam SAX2 coupled with a 2x Navitar microscope objective. A typical experiment duration was $\sim 5$ ms and the camera resolution was set to $768 \times 328$ pixels$^2$ at a sample rate of $30$k frames per second with an exposure time of $2.5$ $\mu$s.

\subsection*{Numerical model}
We consider a liquid jet impacting a pool of identical liquid with velocity $U_0$. The jet is cylindrical with radius $R_j$ and length $L_j$ and is placed at a distance $S$ between the free surface level of the pool and the tip of the jet. The domain is axisymmetric and filled with ambient gas. The top, right and bottom boundaries have outflow conditions imposed with the pressure as $P = P_{\infty}$, zero normal velocity gradients (top and bottom, $\partial u_z/\partial z = 0$; right, $\partial u_r / \partial r = 0$) and zero shear stresses (top and bottom, $\partial u_r / \partial z = 0$; right, $\partial u_z / \partial r = 0$). Since we are studying jet impact in the micro/millimetre regime, effects of gravity are neglected ($g=0$) as hydrostatic effects are small \citep{miguel2021}.

The governing equations are nondimensionalised with the initial radius of the jet $R_j$ and the impact velocity of the jet $U_0$. 

\begin{equation}
    \frac{\partial U_i}{\partial X_i} = 0
    \label{divfree}
\end{equation}

\begin{equation}
    \frac{\partial U_i}{\partial t } + U_0 \frac{U_i}{X_j} =  \frac{1}{\hat{\rho}}\left( -\frac{\partial P}{\partial X_i} + \frac{1}{Re}\frac{\partial(2\hat{\mu} D_{ij})}{\partial X_j} + \frac{1}{We}\kappa \delta_s n_i\right)
    \label{ns}
\end{equation}

Which represent conservation of mass and momentum respectively. In here $U_i$ is the velocity vector, $P$ is the pressure, $D_{ij}$ is the viscous stress tensor, $We = \rho U_0^2 R_j / \gamma$ is the Weber number and $Re = \rho_0 U_0 R_j / \mu$. The last term represents capillary effects, where $\kappa$ is the interface curvature. Ensuring that this term is handled at the liquid interface, the characteristic function $\delta_s$ is used. Lastly, $n_i$ is the normal to the interface. The geometric Volume of Fluid (VoF) method is used to track the interfaces, with a VoF tracer $\Phi$ such that, 

\begin{equation}
    \Phi (x) = 
    \begin{cases}
        1, &\text{if } x \in \text{fluid phase} 
        \\
        0, &\text{if } x \in \text{gas phase}
    \end{cases}
\end{equation}

Therefore, the one-fluid approximation is used in the momentum equation (\ref{ns}) by means of the following arithmetic equations:

\begin{equation}
    \hat{A}(\Phi) = \Phi + (1-\Phi) \frac{A_g}{A_l} \quad\forall A \in \{\rho, \mu\}
\end{equation}

\begin{figure}
    \centering
    \includegraphics[width=1\textwidth]{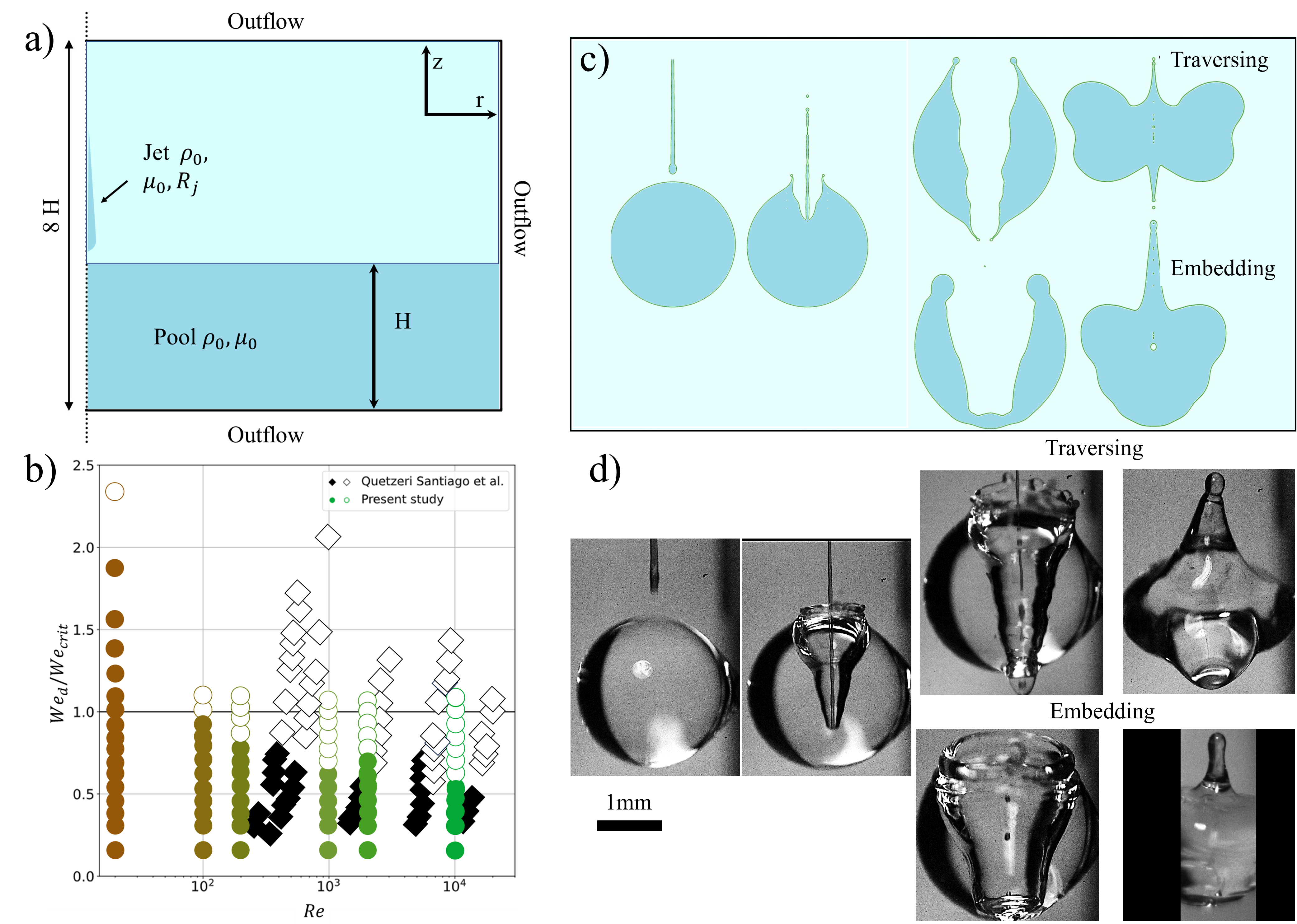}
    \caption{a) Numerical set-up for the study of a jet impact on a droplet. A liquid jet with radius $R_j$ impacts with a velocity $U_0$, viscosity $\mu_0$, and density $\rho_0$ a pool with height $H$ of the same liquid. b) Phase diagram displaying the outcome of droplet penetration based on $Re$ and $We_{jet}/We_{crit}$. With the embedding cases as closed markers and traversing cases as open markers. The experimental data is curved in We-Re space , as it is probed for constant Ohnesorge numbers $Oh = \frac{\sqrt{We}}{Re}$. c) Simulation results of a microfluidic jet impacting a droplet. When the jet has enough inertia to go through the droplet we name it traversing. In contrast if the inertia is not enough we call it embedding. d) Experimental results showing the traversing and embedding of a microfluidic jet on a water droplet \citep{miguel2021}.}
    \label{fig:setup}
\end{figure}

The incompressible Navier-Stokes equations are solved using the finite volume partial differential equation solver Basilisk C \cite{popinet2009, Popinet2018}. With Basilisk, a variety of partial differential equation can be solved with parallelization capabilities on an adaptive mesh refinement (AMR) grid. An example of the mesh refinement used in this work can be seen in figure S1 in the supplementary materials. This solver employs The  Bell-Colella-Glaz (BCG) scheme \citep{Bell1989}, which is a robust second order upwind scheme. In this scheme a projection method is used similar to \cite{Chorin1967} where the pressure and velocity solutions for equations \ref{divfree} and \ref{ns} are decoupled. In this work, we use an improvement on Chorin's method, where we couple the projection and diffusion-convection steps by the BCG scheme. The projection method is also known as a fractional step method, where intermediate iterative steps are used to uncouple the pressure solution while maintaining a divergence-free velocity field.

\subsection*{Validation}
Validation of the code was performed first qualitatively comparing simulations of a microfluidic jet impacting a liquid droplet.  Figures \ref{fig:setup} b) and c), illustrate the capabilities of the numerical technique to reproduce the traversing and embedding phenomena observed in the experiments. The numerical setup is similar to the one in figure \ref{fig:setup}, but instead of a deep pool we initialise a droplet with radius $R_d$. Next, we tested the ability of the code to reproduce the traversing and embedding threshold obtained experimentally and reported in previous works \citep{miguel2021}. After the impact of a microfluidic jet onto a droplet a cavity is created and if the impact velocity is enough to overcome the surface tension of the droplet will traverse it completely. The critical Weber number for traversing the droplet $We_{crit} \approx 64\left(\frac{D_{drop}}{2 D_{jet}}\right)^{1/2}$ was found by comparing the Young-Laplace and dynamic pressures in the cavity. To assess the validity of $We_{crit}$ was experimentally compared with the Weber number based on the jet inertia and the droplet surface tension $\gamma_d$, i.e., $We_d = \rho_0 U_0^2 R_j/\gamma_d$, for a given Ohnesorge number $Oh = \frac{\sqrt{We}}{Re}$. In our simulations, we maintain a constant $Oh$ while varying $Re$, an unattainable condition to the experiments due to the inherent properties of liquids. Figure \ref{fig:setup} b) shows excellent agreement between the experiments and simulations. Furthermore, simulations show that at $Re < 200$ the threshold increases and deviates from the experimental threshold which is lower than the prediction $We_d/We_{crit} = 1$. This indicates that viscous dissipation can influence the traversing process for more viscous liquids than the ones used on the experiments. 

To quantify the numerical convergence, the energy distribution over time is calculated. The supplementary materials provides further details of the energy calculation. We show the energy allocation for different resolutions over the penetration time frame in a bar plot presented in Figure S2 in the supplementary materials. The energy is normalised by the total energy initially present at highest refinement ($r_0 / \Delta = 1024$). From this bar plot we draw multiple conclusions. First we note that over time the total energy is not fully conserved, albeit that increasing the refinement does mitigate the losses. Therefore, we attribute this energy loss to be inherent to the numerical method. Regarding the distribution of energy the fractions are comparable, especially for the three highest refinements. This makes evident that the numerical process converges at resolution $(r_0 / \Delta = 512)$.

\section{Experimental results}

\begin{figure}
    \centering
    \includegraphics[width=1\textwidth]{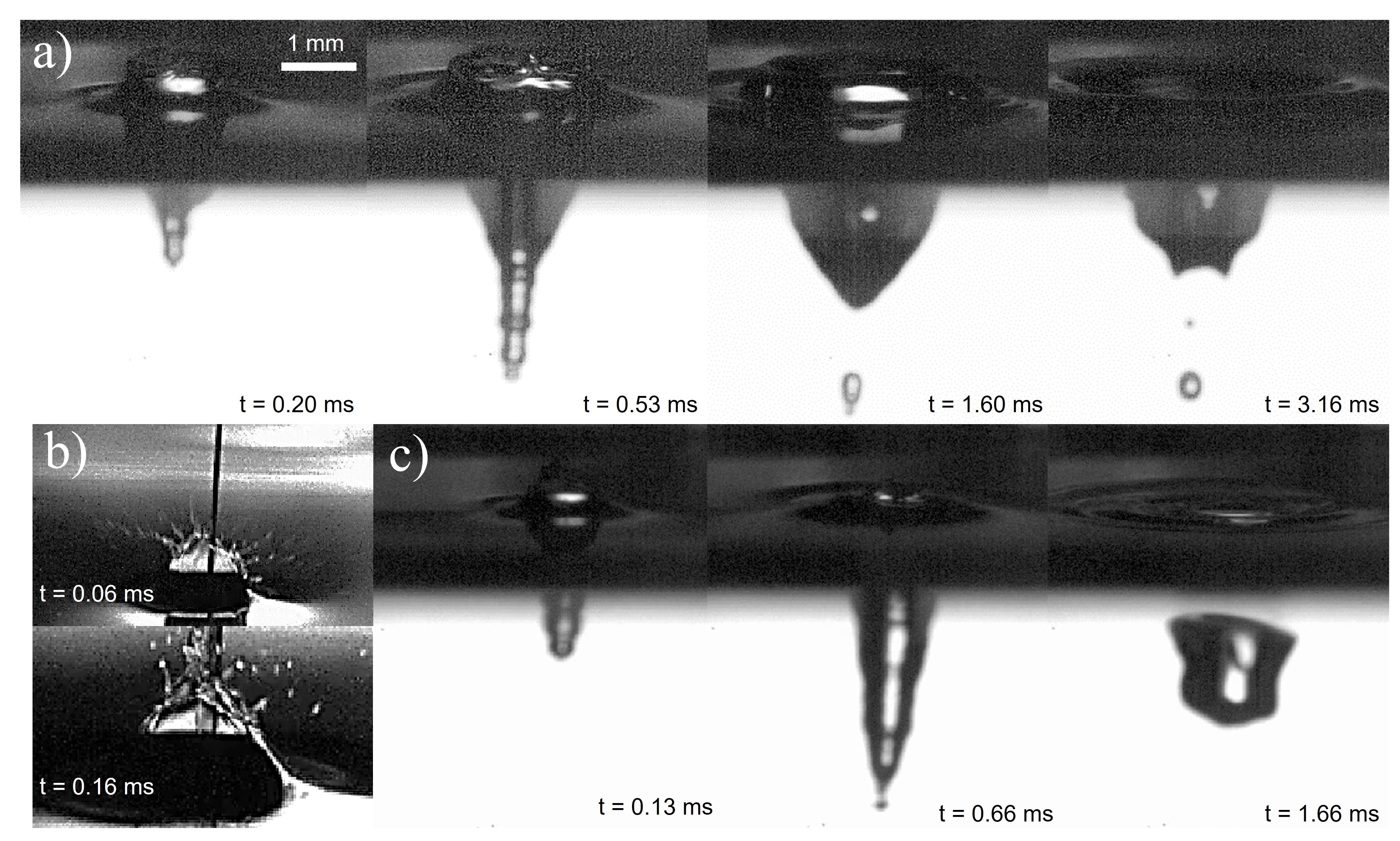}
    \caption{Snapshots of a liquid jet impacting a liquid pool. a) Deep seal, $We = 200$, the cavity collapses below the original position of the pool surface. b) Close up of the surface at the onset the surface seal, $We = 400$. A crown is formed before it collapses on itself. c) Cavity evolution during surface seal. The cavity forms in a similar way to the deep seal case for $t< 0.6$ ms, but afterwards the whole cavity volume remains trapped as a bubble inside the liquid pool.}
    \label{fig:Experiments}
\end{figure}

Similarly to the case of the impact of a microfluidic jet onto a droplet, air is entrained when the jet impacts a deep liquid pool, and a cavity is formed (figure \ref{fig:Experiments}). The cavity continues expanding in both the radial and the $z$ direction, until it collapses.  Previous research shows that for $We >> 1$ during the cavity expansion the process is inertial and the cavity adopts a slender shape \citep{bouwhuis2016}.  Upon reaching the maximum cavity size, interfacial tension starts to influence the cavity dynamics, as kinetic energy is converted into surface energy of the newly formed cavity. The time it takes to reach this regime is approximated by relating the dynamic pressure and the Young-Laplace pressure of the cavity \citep{miguel2021}. 
However, depending on the Weber number the cavity can collapse from the surface (surface seal, figure \ref{fig:Experiments} a) or generate collapse below the original position of the surface (deep seal, figure \ref{fig:Experiments} b). The shape of the cavities and bubbles entrapped are similar to those of impacts on capillary bridges \citep{quetzeri2023cavity}. In these experiments we observe deep seal from $We \approx 35 - 200$, while surface seal is observed from $We \approx 300-400$ (see figure \ref{fig:Experiments}). These findings align with the regime map described in \cite{van2023microfluidic}, categorising them within the ``splashing substrate" region, specifically located at its leftmost boundary, considering a shear modulus G of water equal to 0.

\begin{figure}
    \centering
    \includegraphics[width=1\textwidth]{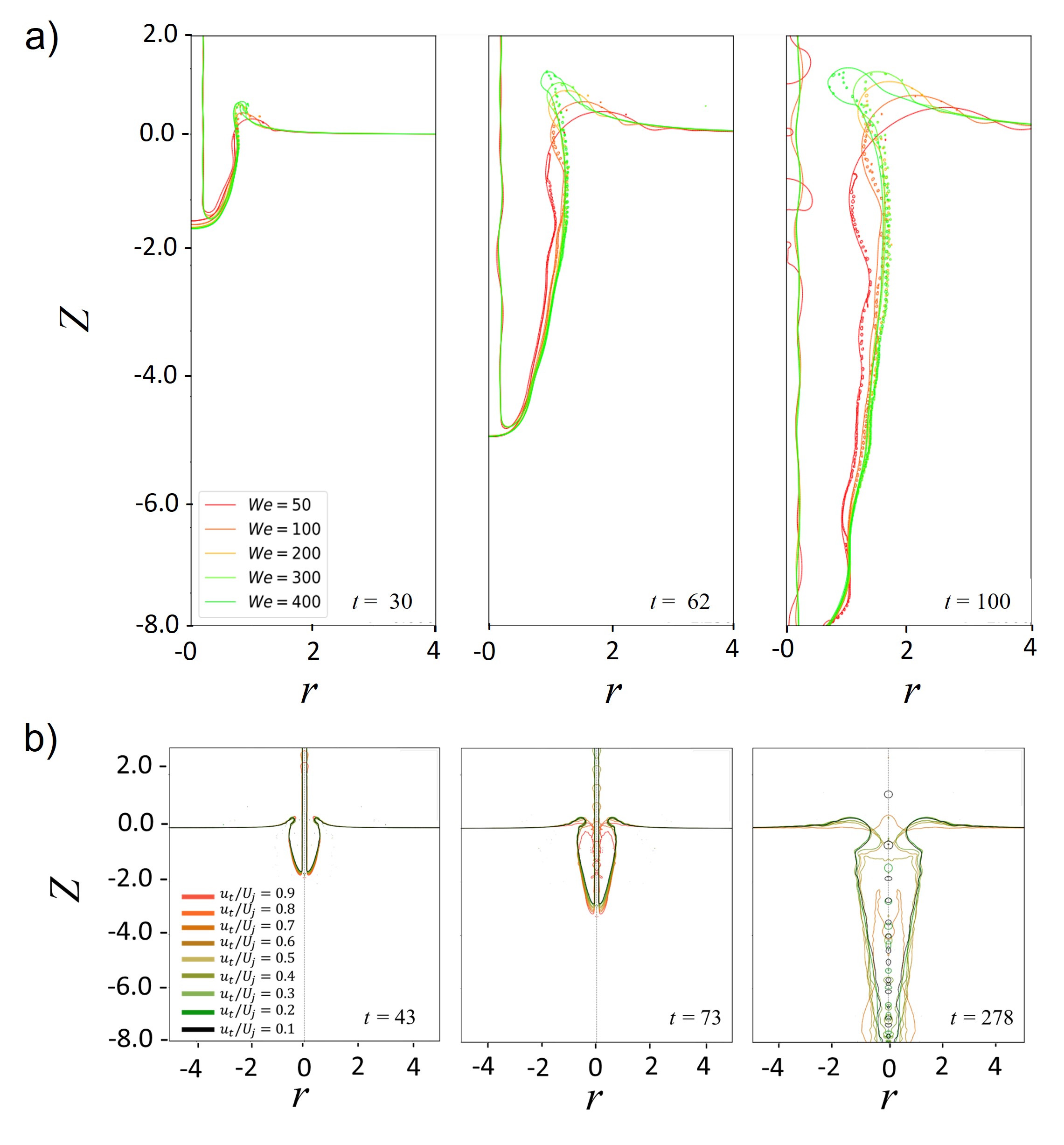}
    \caption{a) Superposition of cavity profiles by cylindrical jets at $Re = 2\cdot 10^4$, for different Weber numbers, indicating inverse relation between rim thickness and Weber number. b) Simulation of jets with $We$ = 200, all starting with identical tip velocities but varying tail velocities. At times $t < 43$, the cavity is similar for all cases. Yet, at time $t = 73$ the cavity collapsed for the cases of $u_t/U_j > 0.8$. At a time $t = 492$, all cavities collapsed with a surface seal, but the cases of $u_t/U_j \leq 0.2$. In general, jets with higher tail velocities exhibit earlier cavity collapse.}
    \label{fig:CollapseProfile}
\end{figure}

In figure \ref{fig:CollapseProfile} a) we show simulation results of the cavity profile evolution of the impact of a cylindrical jet with uniform velocity onto a pool for $We = [50 - 400]$. For all the Weber numbers at $t = 0.6$ the cavity evolution is similar, as is inertia dominated. However, at $t > 1.25$ a deviation from the profiles is observed. Similar to the experiments, for $We > 200$ a rim forms and propels the pool surface upward from its equilibrium surface level. The rim is thinner and shoots higher up as the Weber number increases. In contrast for $We < 100$, surface tension prohibits a slender rim to develop and to advance above the equilibrium surface level. Now the rim of the cavity is flattened and develops into a spherical blob of liquid. Consequently, the seal mechanism differs in both cases, while for $We > 200$ the cavity closure is above the pool equilibrium surface level, the opposite is true for $We < 100$. Here, although the qualitative phenomena is similar, the critical Weber number to transition from surface seal to capillary collapse is shifted by $\approx 30 \%$. Furthermore, an upward jet resulting from the cavity collapse, i.e., a Worthington jet observed in experiments for $We < 200$ is not reproduced in the simulations. 

In the experiments, due to the decelerating nature of the bubble expansion, the impacting jet exhibited a difference between the jet tail velocity $u_t$ and the jet tip velocity $U_j$. Conversely, in our simulations, the velocity remained uniform throughout the entire jet. To bridge this gap, we conducted simulations in which, for simplicity, we implemented a linearly decreasing velocity gradient from the jet's tip to its tail. In this approach, we established the velocity at the tip of the jet as our reference point and systematically adjusted the tail's velocity to 10$\%$, 20$\%$, and so on, up to 100$\%$ of the tip's velocity.

The outcome of these simulations at a $We = 200$ imposing the aforementioned velocity profiles, are depicted in Figure \ref{fig:CollapseProfile}b). We note that a higher tip velocity correlated with an earlier cavity collapse, and the location of cavity collapse approached the surface as $U_t$ increased. Notably, when tail velocities ranged from 0.9 to 0.4 relative to the tip, the impact result manifested as a surface seal. In contrast, tail velocities in the range of 0.3 to 0.1 yielded a deep seal. Consequently, by incorporating a falling linear velocity gradient within our jet simulations, we not only achieved qualitative alignment with our experimental data but also quantitatively replicated the transition from deep to surface seal. However, for simplicity in the remaining of the discussion we keep the jets with uniform velocity distribution.



\section{Cavity dynamics model and simulations}

When the cavity is formed, part of the kinetic energy of the jet transforms in surface energy by the creation of new surface. Thus, the free surface of the cavity has more surface energy than a pool in equilibrium. It is therefore energetically favourable for the interface to restore its rest state. In this way surface tension forces ($F_\gamma$) counteract the radial expansion of a cavity induced by inertia. Nevertheless, jet inertia can also cause a pressure gradient that creates a force that points to the impact centre. By using Bernoulli's principle, along a streamline extending along the surface, one notes that the gas density and velocity contribute to a pressure difference,

\begin{eqnarray}
\left[\frac{1}{2}\rho_g u_g^2 + P\right]_c &=& \left[\frac{1}{2}\rho_g u_g^2 + P\right]_{\infty} \\
 P_{\infty} - P_c &\approx& \frac{1}{2}\rho_g u_g^2 
\label{bernoullideltaP}
\end{eqnarray}

This gradient in pressure induces a force that pulls towards the centre where the pressure is lower. In the remainder of this work, we call this the Bernoulli suction force ($F_{\Delta P}$).

\begin{figure}
    \centering
    \includegraphics[width=0.5\textwidth]{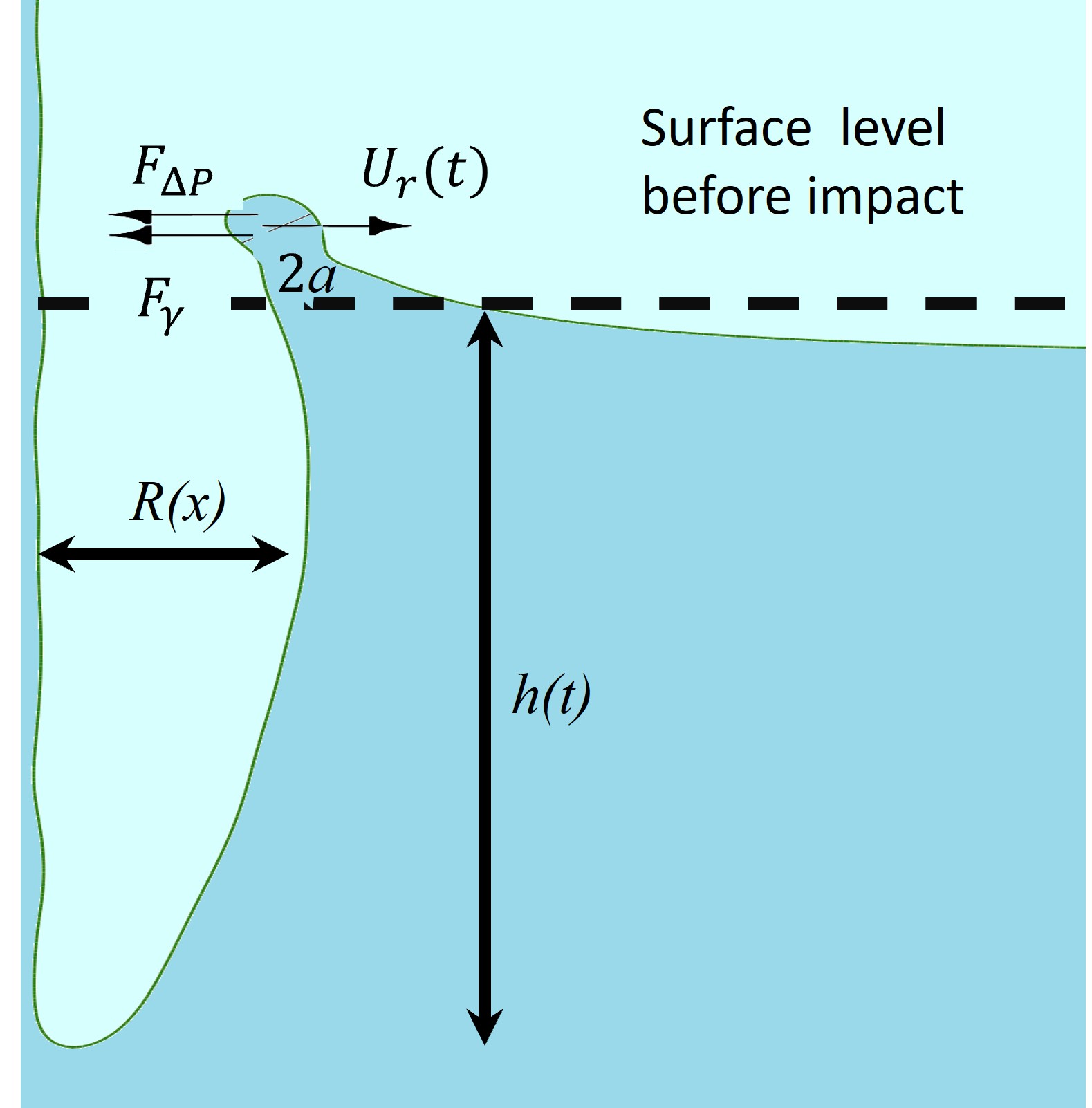}
    \caption{Force diagram on the rim of the cavity and cavity parameters. The rim has a diameter 2a and the forces acting to close the surface are the pressure gradient $F_{\delta P}$ and the surface tension force $F_\gamma$.}
    \label{fig:forcediagram}
\end{figure}

To characterise the collapse time we model the trajectory of the rim of the cavity by considering the radial component of the surface tension force and Bernoulli suction force (see figure \ref{fig:forcediagram}). We assume that the forces only act radially. This enables us to to find analytical expressions for the pinch-off time. However, in reality this is a simplification as it does not consider the rim to translate vertically.

The differential mass of the rim $m = \rho \pi a^2 r(t)d\phi$, where we assume the rim to be circular in cross section, is subjected to the two forces,

\begin{eqnarray}
\label{forcebalance}
m \ddot{r} = F_{\Delta P} + F_{\gamma} &=& -2 a r(t) d\phi \Delta P(t) -4\gamma a d\phi \\ 
\ddot{r} &=& -\frac{2 \Delta P}{\rho \pi  a} - \frac{4\gamma}{\rho a r(t)}
\end{eqnarray}

To get an analytical solution for the radial coordinate of the rim, we look into the limits where one force is negligible, which we explore in the next sections.

\subsection{Radial surface tension regime}
To explain the pinch-off time for $We \sim 1$, we neglect the Bernoulli suction force $F_{\Delta P}$ and assume the only force driving the collapse is the surface tension acting in the horizontal coordinate. Therefore, this problem reduces to that of the collapse of a liquid ring. Since the Reynolds number $Re  >> 1$ , viscous dissipation can be neglected and we can use potential flow to describe the dynamics. Using mass conservation and that the pressure is governed by the Laplace law,  we arrive to the following equation for the evolution of R \citep{texier2013inertial},

\begin{equation}
    \frac{d (R \dot R)}{dt} ln \left(\frac{R+a}{R}\right) + \frac{\dot R^2}{2} \left(\frac{R^2}{(R + a) ^2} - 1\right) = -\frac{\gamma}{\rho R}\left( 1 + \frac{R}{R+a}\right)  
    \label{fulleq}
\end{equation}

Assuming that the thickness of the ring is constant, due to volume conservation $Ra$ is also constant \citep{texier2013inertial}. Therefore, we can linearise equation \ref{fulleq}, considering that $R>>a$ during most time of the closure, and we obtain,

\begin{equation}
    \ddot{R} = - \frac{2 \gamma}{\rho R_0 a_0}.
    \label{linearised}
\end{equation}

By integrating \ref{linearised} and using that at $t = 0$, $\dot R = 0$ and $R = R_0$ we get,

\begin{equation}
    R(t) = R_0 - \frac{\gamma}{\rho R_0 a_0} t^2.
\end{equation}

Therefore, the closure time is 

\begin{equation}
    t_c  = \sqrt{\frac{\rho R_0 a_0}{\gamma}} \sim We
    \label{capclosure}.
\end{equation}

\subsection{Bernoulli suction regime}

For $We >> 1$,  the contribution of surface tension to the collapse can be neglected, and is dominated by the Bernoulli suction force, therefore,

\begin{eqnarray}
\ddot{r} &=& -\frac{2 \Delta P}{\rho \pi  a} = -c_1 \\ 
r(t) &=& -\frac{1}{2}c_1 t^2 + \dot{r_0} t + r_0\\
r(t=t_{c}) &=& -\frac{1}{2}c_1 t_{c}^2 + \dot{r_0} t_{c} + r_0 = 0
\label{quadratic}
\end{eqnarray}

Where as \cite{marston2016} argued, the sheet radius $a$ scales with the surface tension coefficient when applying Taylor-Culick law, which relates the radial speed of holes opening up in a sheet $V_{tc}$ to its thickness and interfacial tension, $a = \gamma/(\rho V_{tc}^2)$. 

Solving equation \ref{quadratic} we get the time for cavity pinch-off,

\begin{equation}
t_{c} = \frac{\dot{r_0}\pm \sqrt{\dot{r_0}^2 + 2 c_1 r_0}}{c_1} \sim \frac{\dot{r_0} a}{u_g^2}.
\end{equation}

We note that at $We >> 1$, the initial velocity of expansion is independent of $We$, i.e., $\dot{r_0} \sim u_g \not\sim f(We)$. Using this fact and Taylor-Culick law for $We >> 1$ the time of closure is, 

\begin{equation}
t_{c} \sim We^{-1}.
\label{airclosure}
\end{equation}


\subsection{Model comparison with simulations}
\begin{figure}
    \centering
    \includegraphics[width=1\textwidth]{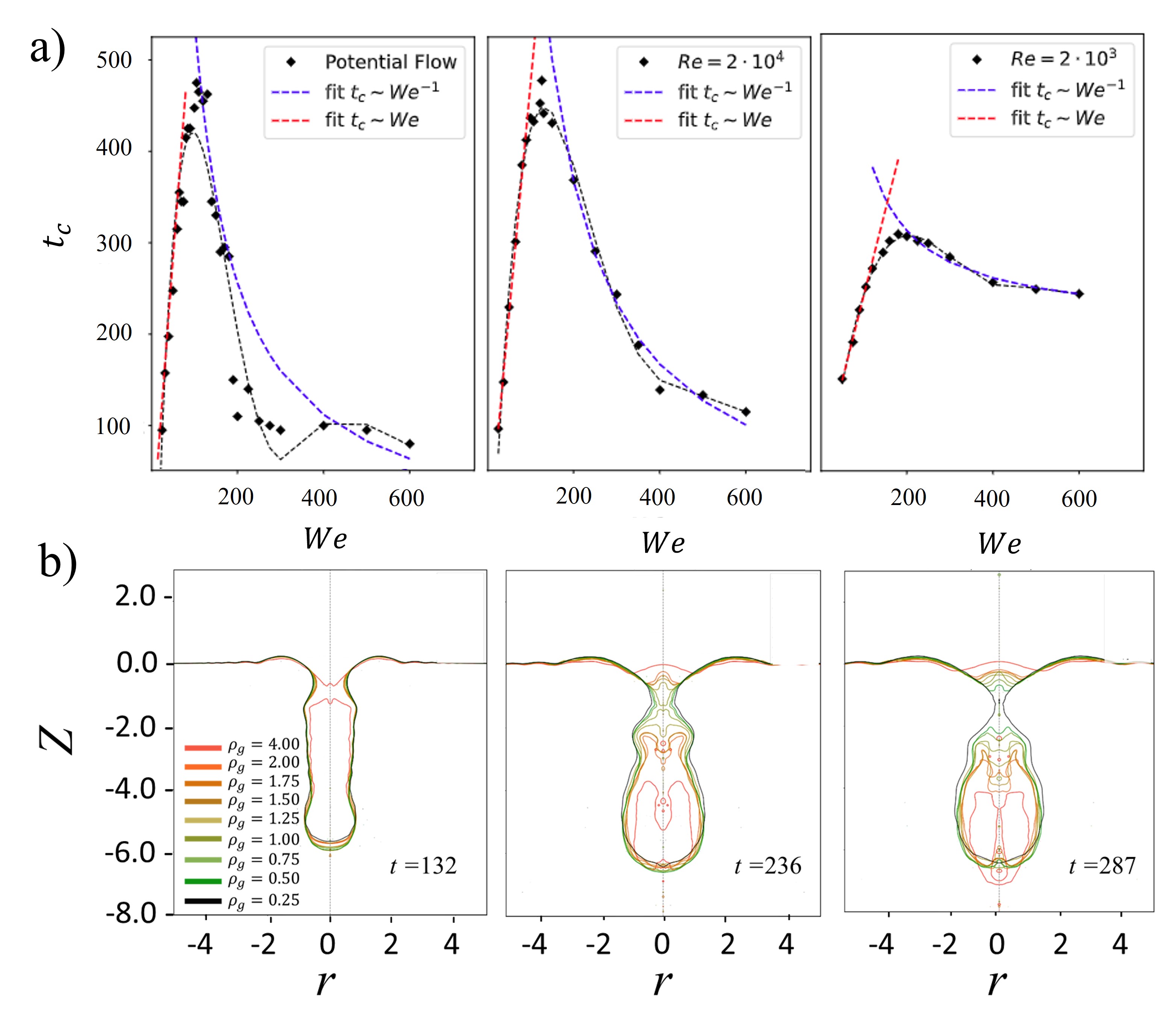}
    \caption{a)Dependency of the time of closure $t_c$ on the Weber number $We$, for different Reynolds numbers $Re$. The black diamonds correspond to the simulation data points. The dotted black line correspond to a spline through the simulation data. The red and blue dotted lines correspond to the approximations given by equations  \ref{capclosure} and \ref{airclosure}. The general trend is similar for each Reynolds number, we observe a global maximum for the time of closure where the two regimes meet. b) Superposition of cavity profiles by cylindrical jets at $Re = 5\cdot 10^3$ and $We = 50$, varying the ambient gas density $\rho_g $ in the range of four times and a quarter to that of atmospheric air. Here we observe that the time of cavity collapse, decreases with increasing air density.}
    \label{fig:closuretime}
\end{figure}

Figure \ref{fig:closuretime} shows the collapse time in terms of the Weber number for four different Weber numbers. Here we observe that in all the cases the closure time reaches a maximum and then it decreases. This can be explained by a collapse regime transition from capillary dominated to the air suction dominated. Indeed, in figure \ref{fig:closuretime} a) we show that the scalings obtained in equations \ref{capclosure} and \ref{airclosure}, match very well with the simulations.  Here we also notice that for a Reynolds number $Re = 2 \times 10^3$, the maximum time of closure is $\approx$ 40 $\%$ than for the potential flow case. This can be attributed to a smaller radial extension due to viscous dissipation. In contrast, for $We = 600$, the time of collapse is $\approx$ 50 $\%$ larger than for the potential flow. Since equations \ref{airclosure} and \ref{capclosure} do not depend on $Re$, we expect that the transition from a capillary collapse to the pressure driven collapse do not depend on the $Re$. Figure \ref{fig:closuretime} shows that the transition occurs at $We \approx  180$ for all the explored $Re$, confirming the closure time independence from $Re$.

\subsection{Effect on ambient gas density}
From our discussion in section 4.2, we would expect that the variation in pressure gradient would determine the time of collapse. Given that the pressure gradient is dominated by the ambient gas density, we expect that an increase in gas density decreases the time of cavity collapse. For $We = 50$, an increase in four times the gas density with respect to the ambient pressure $\rho_g = 4$, results in a cavity collapse at $t \approx 131$, as shown by figure \ref{fig:closuretime}b. In contrast, for gas densities at ambient pressure $\rho_g = 1$, and under ambient pressure $\rho_g = 0.5$, the cavity collapses at $t \approx 236$ and $t \approx 266$ respectively. These results are in line with research on water entry of a sphere, where it was found that the most important parameter gas parameter influencing the lamella ejection is the gas density. Although, in the latter, air density prevents the cavity collapse by resisting the contact line movement \citep{williams2022effect}.

\section{Conclusions}
We investigated the dynamics of high-speed microfluidic jet impacts on a liquid pool, focusing on the formation and collapse of cavities in the micrometer regime. While previous research primarily explored millimetre-scale projectiles, we delved into the range of micrometer-sized projectiles, where surface tension forces dominate the dynamics and hydrostatic pressure can be neglected.

Our experimental setup involved generating high-speed jets from thermocavitation. The resulting jets impacted a water pool, enabling us to qualitatively and quantitatively examine the cavity formation and closure in a parameter range that has not been previously explored. The numerical simulations we employed provided us with the freedom to explore a wide range of parameter combinations, unconstrained by experimental limitations.

Comparing our experimental and numerical results, we observed two distinctive regimes of cavity closure: capillary and air suction-driven. In the capillary-dominated regime $We \leq 150$, surface tension played a predominant role, and we obtained analytical expressions that indicated a closure time scaling with We (equation \ref{capclosure}). In the air suction regime $We \leq 200$, the initial velocity of expansion was the crucial factor, and the closure time scaled inversely with We (equation \ref{airclosure}).

Our findings revealed that the transition from capillary-driven to air suction-driven closure occurred around $We \approx 180$ for the range of Reynolds numbers considered. This transition point was independent of Reynolds number, indicating an inertial dominated phenomena, where viscous dissipation is negligible.  Furthermore, our results shed light on the intricate interplay between gas density and cavity collapse dynamics. These insights into micrometer-scale cavity formation and closure offer valuable knowledge for applications like 3D printing and needle-free injections, and pollutant distribution transport.

\section*{Acknowledgements}
This research was funded by the European Research Council (ERC) under the European Union Horizon 2020 Research and Innovation Programme (grant agreement no. 851630).  M.A.Q-S. acknowledges support from DGAPA through Subprograma de Incorporación de Jóvenes Académicos de Carrera (SIJA). We thank Ulisses Gutiérrez-Hernández for the valuable discussions.

\section*{Declaration of interests}
The authors report no conflict of interest.

\end{document}


\maketitle

\section*{Mesh Refinement Example}

\begin{figure}
    \centering
    \includegraphics[width=0.3\textwidth]{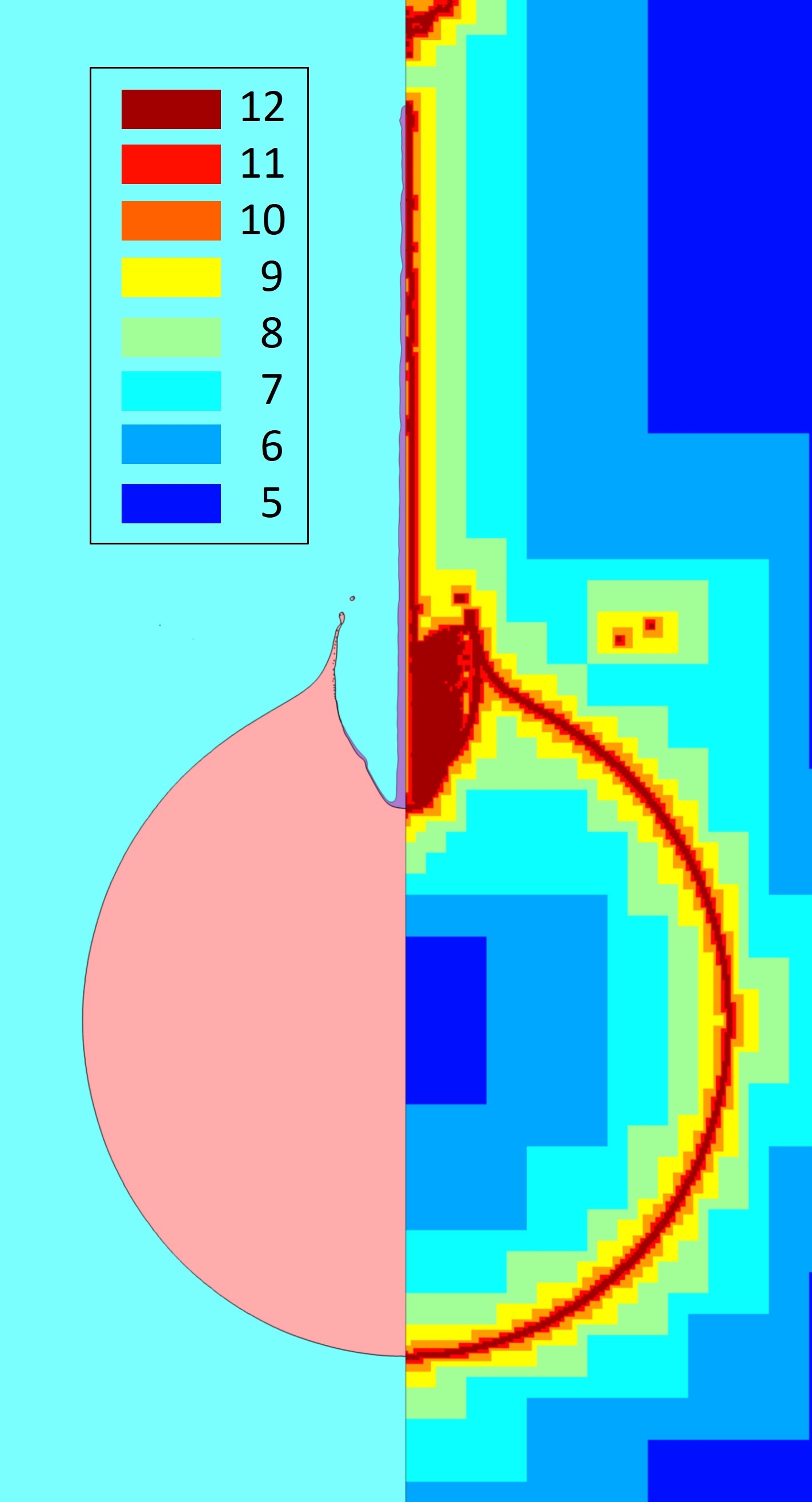}
    \caption{Snapshot of simulated domain for $We = 200$, $Re = 2\cdot10^4$ and $r_0/\Delta \sim 512$. The left side shows the fraction fields of the droplet, jet and gas in pink, purple and light blue, respectively. The right side displays a colour map of the local refinement. The refinement ranges from level 5 to 12, or $1 < r_0 / \Delta < 512$, with decreasing mesh size from blue to red.}
    \label{fig:typsim}
\end{figure}

An instant of the simulation domain is rendered in Figure \ref{fig:typsim}. In this particular snapshot the incoming jet is penetrating the stationary droplet (in pink). The right side gives an insight in the adaptive refinement based on errors in parameters of interest. In this case, the adaptation is done iteratively based on convergence error calculated in the interface curvature and momentum. Note that along the interfaces and around the impacting jet in Figure \ref{fig:typsim} the colour map shows the maximum refinement along the curved interfaces and the jet. 

\section*{Energy Convergence}

\begin{figure}
    \centering
    \includegraphics[width=1.0\textwidth]{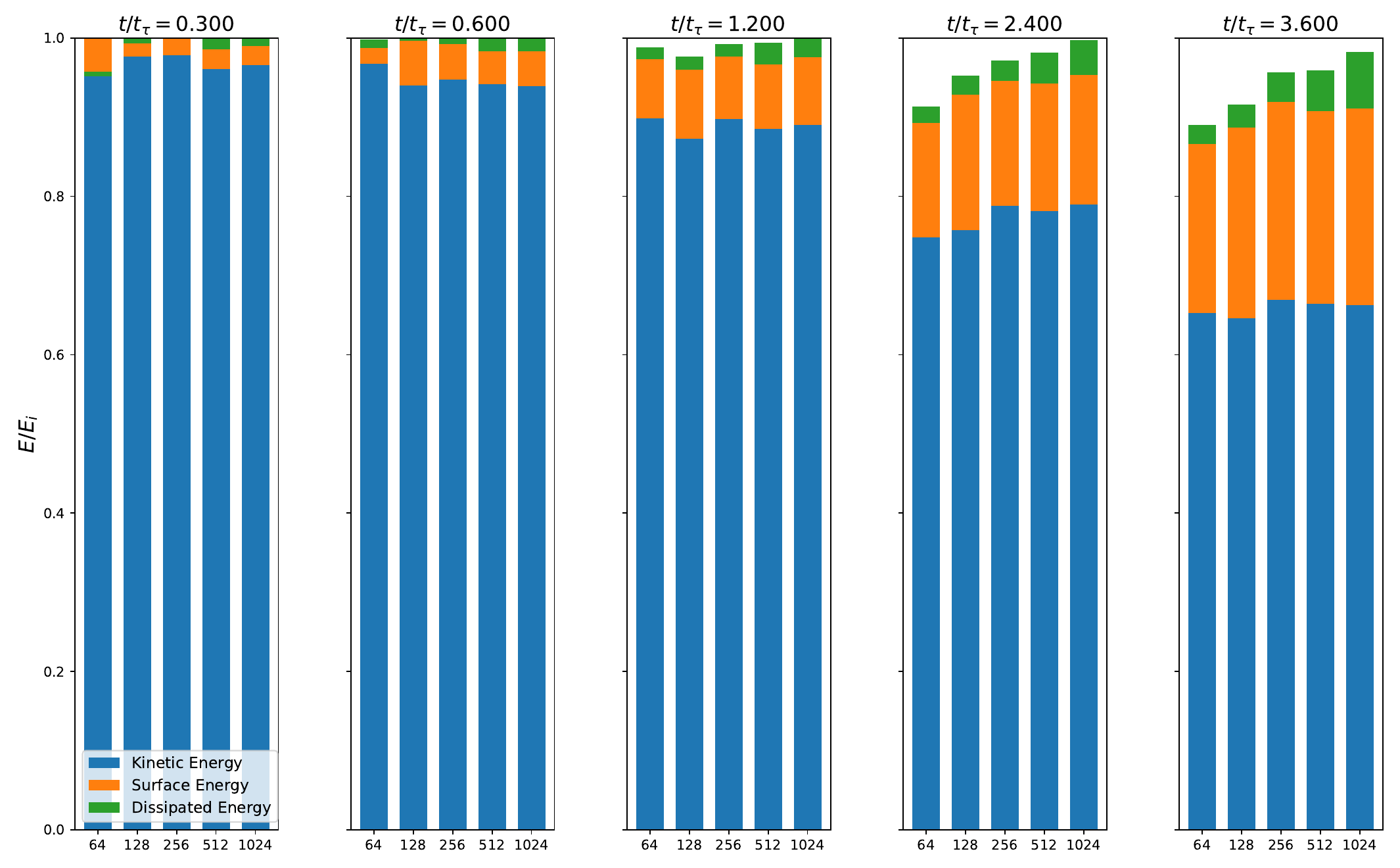}
    \caption{Stacked bar plot showing energy distribution per level of refinement. The energy allocation is depicted for various resolutions across the penetration time frame, with normalisation to the total energy initially present at the highest refinement level ($r_0 / \Delta = 1024$). Total energy is not perfectly conserved over time, with higher refinement level mitigating this loss.}
    \label{fig:GIS_en}
\end{figure}

To quantify the visual numerical convergence, we calculate the energy distribution over time. Figure \ref{fig:GIS_en} shows the energy allocation for different resolutions over the penetration time frame. The energy is normalised by the total energy initially present at highest refinement ($r_0 / \Delta = 1024$). From this bar plot we draw multiple conclusions. First we note that over time the total energy is not fully conserved, albeit that increasing the refinement does mitigate the losses. Therefore, we attribute this energy loss to be an inherent part of the numerical method. Regarding the distribution of energy the fractions are comparable, especially for the three highest refinements. This makes evident that the numerical process converges at resolution $(r_0 / \Delta = 512)$. 

The energy calculation was performed as follows. The total energy consists of the total kinetic energy $E_k$, the total surface energy $E_s$ and the energy dissipation $E_d$:

\begin{equation}
E_k = \frac{1}{2}\iiint_V \left(\hat{\rho} |u_i|^2 \right) \,dV
\end{equation}

In the equation above $\hat{\rho}$ is the arithmetic equation for the liquid density of a particular grid cell, taking a value by means of the expression in equation 2.3 in the main paper.

\begin{equation}
E_s = \iint_{\partial V}\gamma \,dS
\end{equation}

In the above equation, $\gamma$, the surface tension coefficient of the interface present in a particular grid cell, takes the value $r_j/We$. As  $We = \frac{\rho_j u_j^2 r_j}{\gamma}$ where the density and downward velocity of the jet, $\rho_j$ and $u_j$ are 1 by definition.

\begin{equation}
E_d = \int_{t_0}^t \epsilon_{\mu} \,dt
\end{equation}
Where $\epsilon_{\mu}$ stands for the rate of dissipation at a particular instance:
\begin{equation}
    \epsilon_{\mu} = \iiint_V \left(2 \hat{\mu} Re^{-1} |D_{ij}|^2 \right) \,dV
\end{equation}

With $\hat{\mu}$ the arithmetic equation for the viscosity of a particular grid cell and $|D_{ij}|$ denotes the deformation tensor. 

